# Guiding polaritonic energy and momentum through two-dimensional Bravais lattices

Zhonglin Li[1,2], Yingying Wang[1,*], Ruitong Bie[3], Dongliang Yang[3], Tianze Yu[3], Wenjun Liu[1], Linfeng Sun[3,*] and Zexiang Shen[4,*]

[1]Department of Optoelectronic Science, Harbin Institute of Technology at Weihai, Weihai 264209, China

[2]Department of physics, Harbin Institute of Technology, Harbin 150001, China

[3]Centre for Quantum Physics, Key Laboratory of Advanced Optoelectronic Quantum Architecture and Measurement (MOE), School of Physics, Beijing Institute of Technology, Beijing 100081, China

[4]Division of Physics and Applied Physics, School of Physical and Mathematical Sciences, Nanyang Technological University, Singapore 637616, Singapore

Author to whom correspondences should be addressed:

yywang@hitwh.edu.cn; sunlinfeng@bit.edu.cn; zexiang@ntu.edu.sg

**Abstract:**

The strong exciton absorption in monolayer transition metal dichalcogenides provides a promising platform for studying polaritons with tunable dispersions, which are crucial for controlling polaritonic energy and momentum, but remain underexplored. In this work, monolayer $MoS_2$ is coupled with a Fabry-Pérot microcavity to form polaritons. Five types of Bravais lattices with sub-wavelength periods, based on polymethyl methacrylate (PMMA) nanopillars, are intentionally designed. The energy overlap between the periodic PMMA scattering wave and the polariton establishes a coupling channel that controls the directional flow of polaritonic energy, as demonstrated through angle-resolved reflectance measurements. Back-space image measurements further demonstrate that the dispersion in reciprocal space can be directly and manually tuned, allowing for control over their number and their positions. The coupling between the polariton and PMMA scattering wave is further demonstrated by analyzing the reflectance using the two-port two-mode model. The symmetries of 2D Bravais lattices allow the angle between energy and momentum flow to vary widely, from 90°, 60°, 45°, and 30° to arbitrary values. By adjusting the lattice vector lengths, the position of the dispersion branch in a specific direction can be fine-tuned, enabling full-range control over polariton dispersion. This work presents the first theoretical and experimental demonstrations of guiding the direction of polaritonic energy and momentum through Bravais lattice design.

## 1. Introduction

The high exciton binding energy in atomically thin excitonic materials at room temperature provides a compelling platform for creating novel quasiparticles through the hybridization of excitons with various fundamental physical particles within a quantum many-body system. The coupling of excitons in monolayer transition metal dichalcogenides (TMDs) with photons in microcavity and photonic crystal leads to the formation of exciton polaritons, which are half-light, half-matter bosonic quasiparticles. The strong light-matter coupling, along with ultrafast polaritonic carrier dynamics, maximized light trapping, and the absence of a population inversion requirement, promotes the development of low-power polaritonic light sources.[1-6] Additionally, exciton polaritons in adjacent monolayer $MoS_2$ can couple to form interlayer polaritons, which exhibit strong nonlinearity and promote the study of few-polariton quantum correlations.[7] Besides, the inclusion of metal nanostructures in TMDs further enables the formation of hybrid exciton-plasmon-polaritons, which are particularly suitable for generating superlinear light.[8,9] The study of polaritons advances research into various fascinating physical phenomena, including laser emissions,[2] room-temperature polariton spin switches,[10] exciton–polariton condensation,[11] and polariton Bose–Einstein condensates.[12] Recently, polaritonic qubits have been generated for the realization of quantum gates and algorithms, analogous to quantum computation with standard qubits.[13,14]

Flexibly creating and manipulating exciton polaritonic states with specific energy and momentum is crucial for applications in photonics and quantum technologies.[15,16]

The tunable momentum of polaritons introduces more allowed scattering processes in reciprocal space, in which polaritons can participate, facilitating the directional control of radiation and other scattering processes. The control of polaritonic dispersion, including the slope of the dispersion and the number of dispersion branches, is significant for designing optical devices with specific properties, such as coupling light of different wavelengths at the same incident angle and controlling light reflectance at arbitrary energy and incident angle. The study of polaritonic dispersion also enhances the understanding of the coupling between polaritonic propagation waves. However, the manipulation of the directional flow of polaritonic energy in monolayer TMDs within reciprocal space remains relatively unexplored.

This issue can be addressed by introducing reciprocal lattices, under the consideration of momentum conservation. The conservation of momentum has long played a significant role in determining diffraction patterns for periodic structures, such as multiple-order diffractions for gratings, or guided resonance modes and bound states in the continuum for reflected and transmitted light.[17-21] Furthermore, recent advances in optical Fourier surfaces based on Ag films have enabled access to previously unattainable diffractive surfaces, including two-dimensional (2D) moiré patterns, quasicrystals, and holograms.[22] Recently, coupling photonic modes with laser gain materials at high-symmetry points beyond the Γ point in reciprocal space has enabled the realization of angle-tunable lasers.[23] In addition, directional polariton lasing at corner states has been realized based on perovskite film, after pattern them into a 2D Su-Schrieffer-Heeger lattice.[4] Therefore, periodic structures with different symmetries provide an efficient platform for tuning the allowed wavevector of scattered light in reciprocal space under the restriction of momentum conservation.

In this work, a Fabry-Pérot (F-P) cavity is constructed based on monolayer $MoS_2$ supported on a $SiO_2$/Si substrate. The coupling of A and B excitons in monolayer $MoS_2$ with photon in microcavity results in the formation of polaritons, whose dispersions including high polariton, middle polariton, and low polariton, are individually revealed by angle-resolved reflectance measurements, with Rabi splittings of 104 meV and 194 meV, respectively. In addition, it is found that the dispersion of the polymethyl methacrylate (PMMA) scattering wave is manually tuned in reciprocal space by translating different reciprocal lattices, after constructing 2D Bravais lattices based on PMMA nanopillars. The energy overlap between the PMMA dispersion and polariton dispersion establishes a coupling channel that controls the directional flow of polaritonic energy, as observed through angle-resolved reflectance measurements. Back-space imaging measurements further demonstrate that the dispersion can be directly and manually tuned, allowing adjustments to both their number and their positions. Additionally, the coupling between the polariton and PMMA surface wave is further demonstrated by analyzing the reflectance by using the two-port two-mode model. The enhanced light-matter interaction, precise control of polariton energy and momentum, and the ability to manipulate the directional flow of polaritonic energy obtained here enrich the study of polariton optics in the monolayer TMDs-coupled nanostructures.

## 2. Results and discussions

## 2.1 The design concept for realizing the directional control of polaritonic energy and momentum based on 2D Bravais lattices

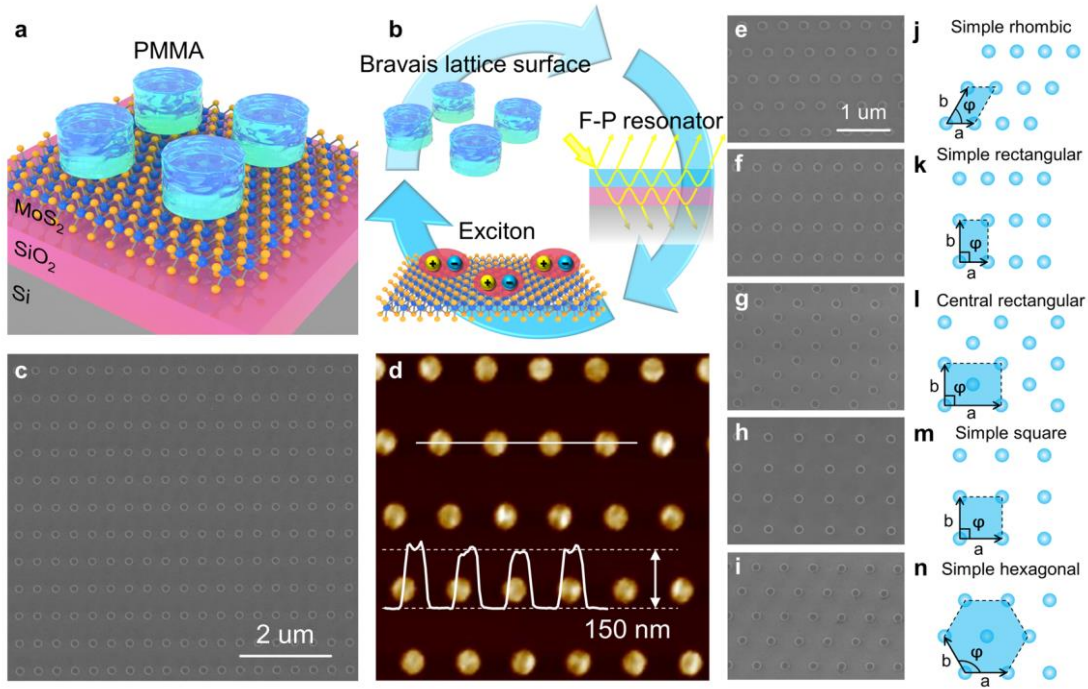

**Fig. 1 2D rhombic Bravais lattice design, fabrication, and characterization. a,** Schematic illustration of a 2D rhombic Bravais lattice based on PMMA nanopillars supported by monolayer $MoS_2/SiO_2$(300 m)/Si. **b,** The design concept for realizing directional control of polaritonic energy and momentum. **c,** Scanning electron microscope (SEM) image of the fabricated large-scale rhombic Bravais lattice, with primitive lattice constants of a=400 nm and b=600 nm, formed by the periodic arrangement of PMMA nanopillars with a diameter of 160 nm and a height of 150 nm. **d,** Atomic force microscope (AFM) image of the fabricated rhombic Bravais lattice, showing a height difference of 150 nm between the PMMA nanopillar and the supporting substrate. **e-i,** and **j-n,** SEM images of five fabricated Bravais lattices based on PMMA nanopillars, along with their schematic representations for their primitive lattices.

**Fig. 1a** shows a schematic illustration of a 2D rhombic Bravais lattice based on PMMA nanopillars supported by monolayer $MoS_2/SiO_2$(300 m)/Si. An F-P cavity is constructed using high-reflective $MoS_2$ and Si, with $SiO_2$ serving as the space layer. The coupling of the PMMA scattering wave, exciton, and microcavity photons simultaneously is possible in this structure, as seen in **Fig. 1b**. Monolayer $MoS_2$ is grown on a $SiO_2$(300 nm)/Si substrate by chemical vapor deposition (CVD). The PMMA layer is spin-coated on the monolayer $MoS_2/SiO_2$/Si, and nanopillars are fabricated by electron beam lithography (EBL) to form different 2D Bravais lattices (Method). **Fig. 1c** gives a SEM image of the fabricated large-scale rhombic Bravais lattice, with primitive lattice constants of a=400 nm and b=600 nm, formed by the periodic arrangement of PMMA nanopillars with a diameter of 160 nm and a height of 150 nm. **Fig. 1d** shows an AFM image of the fabricated rhombic Bravais lattice, showing a height difference of 150 nm between the nanopillar and the substrate. **Figs. 1e-1i** display SEM images of five fabricated Bravais lattices based on PMMA nanopillars, along with their schematic images for their primitive lattices (**Figs. 1j-1n**). Four major crystal systems and five 2D Bravais lattices are systematically studied here. It can be observed that the Bravais lattices exhibit different symmetries, allowing for the simultaneous control of symmetries for reciprocal lattice vectors. The SEM images of these fabricated Bravais lattices and their Fourier transforms are provided in **Table S1** of **Supplementary Note 1**.

**2.2 The formation of polaritons through the coupling of excitons in monolayer $MoS_2$ with photons in an F-P microcavity**

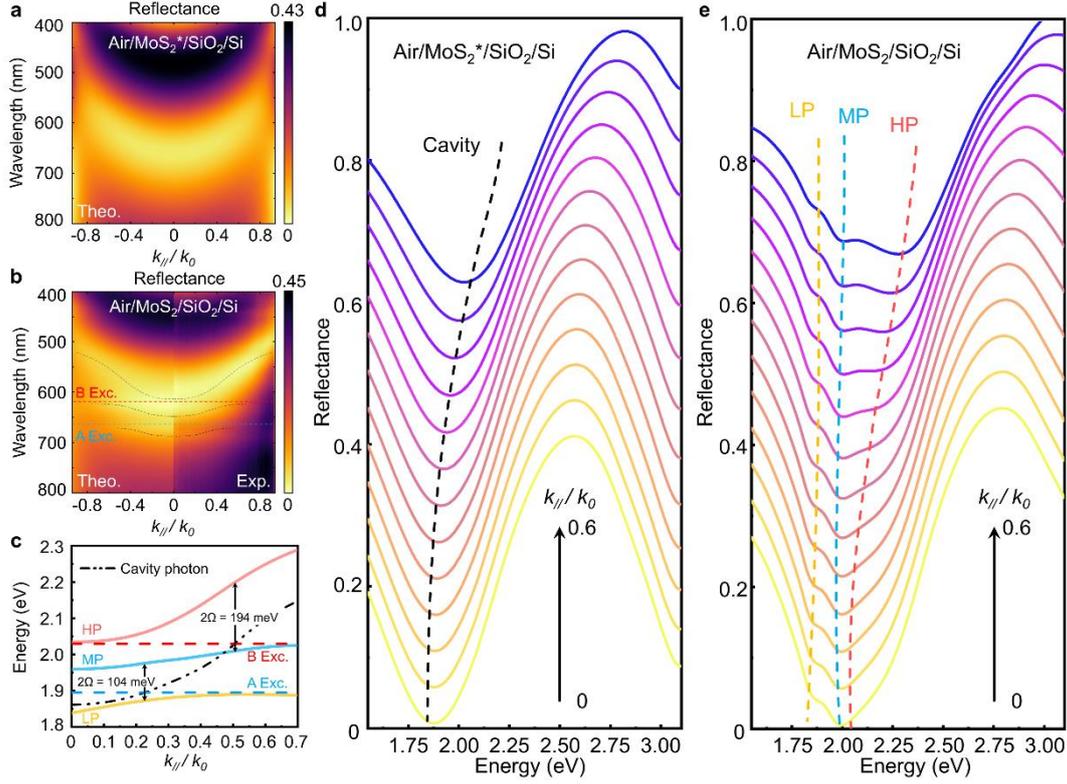

**Fig. 2 Coupling MoS₂ excitons to photons forms polaritons in F-P microcavity. a,** Theoretical angle-resolved reflectance of monolayer MoS$_2$ supported on a SiO$_2$(300 nm)/Si substrate. The A and B exciton absorptions are intentionally excluded from this calculation. **b,** Theoretical and experimental angle-resolved reflectance of monolayer MoS$_2$ supported on a SiO$_2$/Si substrate. The A and B exciton absorptions are included in this calculation, and the formation of exciton polariton dispersion is observed. **c,** The splitting of the photon dispersion into a hybrid upper polariton dispersion, middle polariton dispersion, and lower polariton dispersion, when the cavity photon is resonant with the A and B excitons in MoS$_2$ by tuning the incident angle. Rabi splittings of 0.104 eV and 0.194 eV are observed for the exciton polaritons. **d,e,** The individual reflectance for the cavity at different incident angles, without and with consideration of exciton absorptions of monolayer MoS$_2$.

**Fig. 2a** shows the theoretical reflectance of monolayer MoS$_2$ supported on a SiO$_2$(300 nm)/Si substrate, as a function of $k_{\parallel}/k_0$. Here, $k_{\parallel}/k_0=\sin(\theta)$ is the ratio of the in-plane component ($k_{\parallel}$) of the incident light to the wave vector of the incident light (**k₀**), where θ is the incident angle. The A and B exciton absorptions are intentionally excluded from this calculation. The refractive index for MoS$_2$ without and with exciton absorption is provided in **Fig. S1** of **Supplementary Note 2**. The calculation is based on the transfer matrix method for a multilayer structure. It is observed that there is an extremely low reflectance (a reflection dip) with a broad peak width in the visible light range, indicating the generation of the F-P resonance mode in this multilayer structure. In addition, the F-P resonance mode undergoes a blue shift as the incident angle increases, due to the dispersion of the cavity photon, as shown for the individual reflectance spectrum provided in **Fig. 2d**.

Furthermore, when the cavity photon is resonant with the excitonic transition energy, hybrid polaritons are formed, and anti-crossing behavior is observed in the band structures, as shown in **Figs. 2b-2c**. **Fig. 2b** gives the theoretical and experimental angle-resolved reflectance of monolayer MoS$_2$ supported by a SiO$_2$(300 nm)/Si substrate, with consideration of exciton absorptions. The exciton absorptions in MoS$_2$ are highlighted by red and green dashed lines. It is observed that when the cavity photon is resonant with the exciton, the dispersion of the cavity photon splits into three polaritonic branches, labeled as low polariton (LP), middle polariton (MP), and high polariton (HP), as indicated by black dashed lines in **Fig. 2b**. The Rabi splitting is a measure of the coupling strength between the cavity photon and the exciton, which can be calculated using the coupled oscillator model.[1] Rabi splittings of 0.104 eV and 0.194 eV for exciton-polaritons are obtained, as shown in **Fig. 2c**. **Fig. 2e** gives the individual reflectance of the cavity at different incident angles with exciton absorptions. The dispersion of different polaritons and the anti-crossings between these polaritonic branches are clearly observed.

Therefore, the multilayer structure presented here provides a coupling channel between the microcavity photon and the exciton. The coupling strength between the photon and the exciton alters the slope of dispersion curve, thereby allowing for the tuning of group velocity and the density of state for polaritons. In addition, this F-P cavity serves as a platform for coupling photons with excitons for other excitonic materials. The angle-resolved reflectance of monolayer $WS_2$, $MoSe_2$, and $WSe_2$ is theoretically studied in **Fig. S2** of **Supplementary Note 2**. It is observed that when the dispersion of cavity photon overlaps with exciton absorption, exciton polaritons are formed, causing the dispersion of the F-P modes to split into multiple polaritonic branches.

## 2.3 The dispersion of the periodic PMMA scattering wave is manually tuned using 2D Bravais lattices

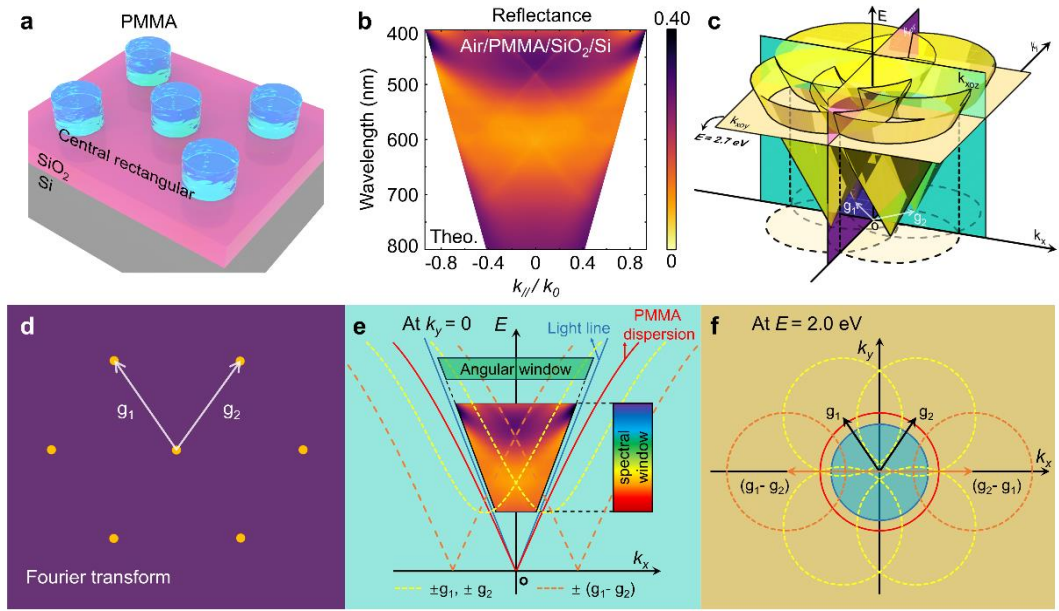

**Fig. 3 Tuning PMMA dispersion with 2D Bravais lattices. a,** Schematic image of a 2D central rectangular Bravais lattice based on PMMA nanopillars supported by a $SiO_2$(300 nm)/Si substrate. **b,** Theoretical mapping of angle-resolved reflectance of this Bravais lattice. **c,** The full dispersion diagram of the periodic PMMA scattering wave, whose dispersion cone is displaced by four principle reciprocal lattice vectors, $\pm\mathbf{g_1}$ and $\pm\mathbf{g_2}$, resulting in four cones. Two slices through the full dispersion diagram are used to probe the PMMA dispersion in reciprocal space. **d,** The Fourier transform of the central rectangular Bravais lattice, with principle reciprocal lattice vectors, $g_1$ and $g_2$, highlighted by arrows. **e,** Schematic image of the dispersion diagram (energy versus in-plane wave vector $k_x$ at $k_y=0$) for the central rectangular Bravais lattice. The blue line is the light line from the light cone. The red line represents the PMMA dispersion, while the yellow dashed lines show the dispersion shift by the principle reciprocal vectors, $\pm g_1$, $\pm g_2$. The orange dashed lines show the dispersion shift by the reciprocal lattice vectors, $\pm(g_1-g_2)$, individually. The trapezoidal region indicates the experimentally accessible area on the dispersion diagram. **f,** Schematic diagram of a slice through the full dispersion diagram (along the $k_x$-$k_y$ plane) in panel (c) at a fixed energy.

**Fig. 3a** shows a schematic image of a 2D central rectangular Bravais lattice based on PMMA nanopillars supported by a $SiO_2$(300 nm)/Si substrate, with its Fourier transform provided in **Fig. 3d**. Two principle reciprocal lattice vectors, $g_1$ and $g_2$, and numerous reciprocal lattice vectors, **g**, can be formed by linear combinations of $g_1$ and $g_2$, as seen from **Fig.3d**. **Fig. 3b** gives the theoretical angle-resolved reflectance of this lattice as a function of $k_\parallel/k_0$ (Method). The individual spectrum is studied in **Fig. S3** of **Supplementary Note 3**. In the visible light range, a reflection dip with a broad peak width is observed, originating from the F-P cavity mode. In addition, multiple dispersion branches are observed, corresponding to multiple reflectance dips with very sharp peak widths. These branches originate from shifts in the PMMA dispersion, as discussed below, and their formations are schematically given in **Fig. 3e**. Analogous to electrons in crystals, where the Bloch wave traveling in a periodic potential result in the formation of an electron band structure. The PMMA scattering wave travels in periodic dielectric environments, resulting in the formation of multiple dispersion branches. The control of number of dispersion branches makes multiple low reflectance at arbitrary energy and incident angle possible. Furthermore, the angular reflectance for PMMA-based Bravais lattices, including the simple rhombic, simple rectangular, simple square, and simple hexagonal lattice, is provided in **Fig. S4** of **Supplementary Note 4**.

In addition, it is further demonstrated that the optical properties of those dispersion branches could be tuned through nanopillar's diameter, height, and dielectric function, as provided in **Fig. S5- Fig. S7** of **Supplementary Note 5**-**7**. It is found that The PMMA nanopillars exhibit dispersion control similar to that of Au or Ag plasmonic nanopillars, including dispersion branches and quality factors. Therefore, the periodic structure of PMMA nanopillars offers a new approach for controlling the dispersion of optical fields. Besides, PMMA is a commonly used photoresist, and the fabrication of its nanostructures is relatively straightforward. However, the fabrication process of Au and Ag nanopillars inevitably involves etching of the metal films, which can introduce defects into the underlying monolayer $MoS_2$. In contrast, the etching process for PMMA is more mature and well-established, making it less prone to defect introduction.

**Fig. 3c** shows the full dispersion diagram of the periodic PMMA in reciprocal space, including multiple dispersion cones. The periodic structure provides momentum to the impinging light, leading to the formation of electromagnetic scattering waves with in-plane wavevector $\mathbf{k}_{sca}$, which propagate in PMMA with an optical dispersion cone.[22] The conservation of momentum allows the generation of additional scattering waves with wavevectors given by $\mathbf{k'}_{sca}=\mathbf{k}_{sca}\pm\mathbf{g}$, where, $\mathbf{k'}_{sca}$ and $\mathbf{k}_{sca}$ are the wavevectors of scattering waves propagating in different directions, and **g** is the reciprocal vectors, under the consideration of elastic scattering. Thus, the dispersion at the Γ point is shifted by amounts of ±**g**$_1$ and ±**g**$_2$, resulting the duplication of one dispersion cone into four, as schematically given in this figure. It should be noted that due to the large number of reciprocal lattices, a large number of dispersion cones can be generated. However, the used measurement equipment limits the dispersion that could be experimentally probed.

The PMMA dispersion can be optically probed in different **k**-planes, such as along the **k**$_x$-**k**$_z$ plane or the **k**$_x$-**k**$_y$ plane, where **k**$_x$, **k**$_y$ and **k**$_z$ are the x, y, and z components of the wavevector, as given in **Figs. 3c** and **3e**. **Fig. 3e** shows the diagram of angle-resolved reflectivity as a function of **k**$_x$. The solid blue line represents the light line from the light cone ($E = \hbar c k$). The red line represents the dispersion of the periodic PMMA scattering wave. When **k**$_x$±**g**=**k**$_{sca}$ is satisfied, a decrease in reflectivity is observed. The yellow dashed lines show the dispersion shift due to the principle reciprocal lattice vector (±**g**$_1$, ±**g**$_2$). The orange dashed lines shows the dispersion shift due to the next nearest-neighbor reciprocal lattice vectors (±(**g**$_1$-**g**$_2$)). The experimentally accessible area on the dispersion diagram is further highlight by a trapezoidal region, which is limited by the spectral window of the spectrometer along energy, and the angular window of the reflected light collected by the microscope objective along **k**$_x$. Based on above discussions, the reflectance for the PMMA-based Bravais lattice can be easily understood by considering the dispersion shift due to reciprocal vectors. The dispersion shift causes the structure to couple light of different wavelengths at the same incident angle, e.g., two reflectance dips at different energies are simultaneously observed at the Γ point, as shown in **Fig. 3e**. Away from the Γ point, additional reflectance dips at different wavelengths are observed.

**Fig. 3f** shows a schematic image of a slice through the dispersion diagram in **Fig. 3c** at a fixed energy. As can be seen, different numbers of arcs can be observed in the k-space image measurement, determined by the spectral window of the equipment used. Each arc pair represents a solution to **k**$_∥$±**g**=**k**$_{sca}$, and different reciprocal lattice vectors bring arc pairs at different positions. Accordingly, the momentum of the scattering wave could be manually tuned in reciprocal space by periodic structure design.

**2.4 The number and position of dispersions can be manually tuned using 2D Bravais lattices**

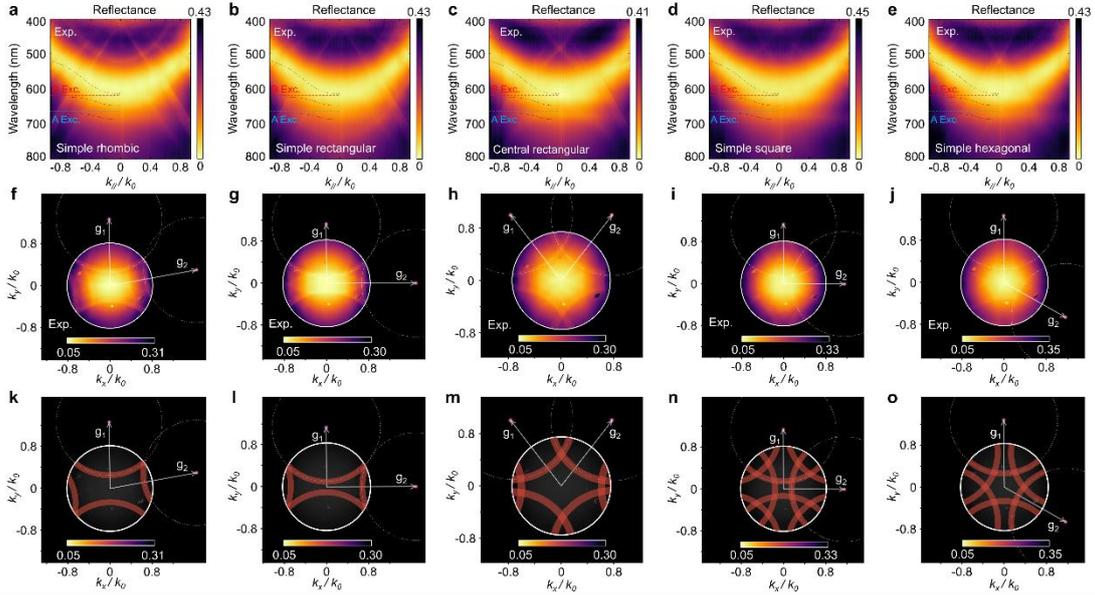

**Fig. 4 Experimental reflectance and k-space images of five 2D PMMA Bravais lattices supported by a monolayer MoS$_2$/SiO$_2$/Si substrate. a-e,** Experimental results of angle-resolved reflectance for five 2D Bravais lattices based on PMMA nanopillars supported by a monolayer MoS$_2$/SiO$_2$(300 nm)/Si substrate. **f-j,** Experimental results of k-space reflectivity images for those five 2D Bravais lattices based on PMMA

nanopillars supported by a monolayer $MoS_2$/$SiO_2$(300 nm)/Si substrate, with 633 nm photons incident on the structures. **k-o,** Arc pairs in the k-space reflectivity images. The reflectance from the middle polariton branch has been removed, in order to observe the dispersion shift due to the periodic structure.

**Figs. 4 a-e** show the experimental results of angle-resolved reflectance for five 2D Bravais lattices based on PMMA nanopillars supported by a monolayer $MoS_2$/$SiO_2$(300 nm)/Si substrate. In the visible light range, the dispersion of F-P microcavity coupled with exciton in $MoS_2$, and resulting in three polaritonic dispersion branches as discussed before. In addition, multiple dispersion branches from PMMA are clearly observed. The dispersion branches of PMMA cover a broad wavelength range and overlap with the polaritonic dispersion, creating a coupling channel between the PMMA scattering wave and polaritons, which directs polaritonic energies in different directions through the PMMA channel.

**Figs. 4f-4j** show the measured k-space reflectivity images for 633 nm photons (1.96 eV, energy selected at middle polaritonic branch) incident on the periodic structures, with $k_x$ and $k_y$ normalized by $k_0$. In these figures, decreased reflectivity from middle polariton dispersion is observed over a large range in **k**-space. In addition, for the five 2D Bravais lattices, four to eight arcs (highlighted by white dashed lines) appear due to decreased reflectivity, when $k_\parallel \pm g = k_{sca}$ is satisfied. To observe the arcs individually, the reflectivity from the polaritons have been removed, as schematically depicted in **Figs. 4k-4o**. It is further demonstrated that the polaritons couple with periodic PMMA, and the polaritonic energy flows into the PMMA, whose periodic

structure further promotes the redistribution of polaritonic energy and momentum in reciprocal space.

Furthermore, the symmetry reduction flowchart for the five 2D Brava lattices is further given in **Fig. S8** of **Supplementary Note 8**. It is found the different symmetries of 2D Bravais lattices enable directional control of photonic dispersion. The angle between the directions of energy and momentum flow can vary from 90°, 60°, 45°, and 30° to arbitrary value. The position of the dispersion branch in a specific direction can be fine-tuned by adjusting the lengths of the primitive lattice vectors. Consequently, 2D Bravais lattices provide full-range control over the dispersion of polaritons.

**2.5 The coupling between polaritons and periodic PMMA scattering waves**

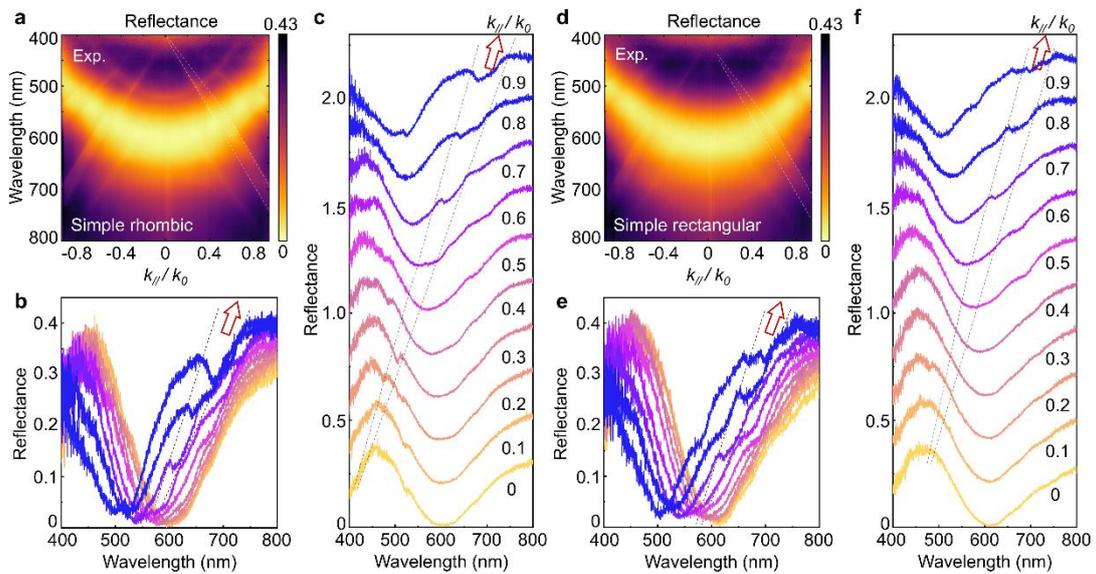

**Fig. 5 Angle-resolved reflectance reveals polaritons coupling with PMMA scattering waves. a,** Experimental results of angle-resolved reflectance for a simple rhombic Bravais lattice based on PMMA nanopillars supported by a monolayer $MoS_2$/$SiO_2$(300 nm)/Si substrate. **b,c,** Experimentally measured angle-resolved reflectance spectra at different incident angles, with a vertical shift introduced in panel (c) for clarity. **d,** Experimental results of angle-resolved reflectance for a simple

rectangular Bravais lattice based on PMMA nanopillars supported by a monolayer MoS$_2$/SiO$_2$(300 nm)/Si substrate. **e,f,** Experimentally measured angle-resolved reflectance spectra at different incident angles, with a vertical shift introduced in panel (f) for clarity.

**Fig. 5a** shows the experimental results of angle-resolved reflectance for a simple rhombic Bravais lattice based on PMMA nanopillars supported by a monolayer MoS$_2$/SiO$_2$(300 nm)/Si substrate. The reflectance dip from the PMMA Bravais lattice in the visible light range is systematically studied, as marked by dashed lines in **Figs. 5a** and **5b**. **Figs. 5b** and **5c** further show the experimentally measured angle-resolved reflectance spectra at different incident angles, with a vertical shift introduced in **Fig. 5c** for clarity.

For PMMA nanopillars supported by the monolayer MoS$_2$/SiO$_2$/Si substrate, the reflection dips are further analyzed using the two-port two-mode resonator system. In this configuration, the MoS$_2$/SiO$_2$/Si F-P cavity can be considered as a separate cavity that can couple with the PMMA cavity. For the PMMA based cavity, both reflection and transmission are allowed. According to the temporal coupled-mode theory, the complex reflection coefficient from the PMMA at frequency around resonance ($\omega_0$) can be written as Equation (1).[24-26]

$$r = -1 + \frac{1/Q_r}{-i(\omega/\omega_0 - 1) + (1/Q_r + 1/Q_a)/2} \tag{1}$$

Here, $\omega$ and $\omega_0$ are the frequencies of the incident light, around and at resonance, individually. This resonator is defined by dissipation-related parameters $Q_r$ and $Q_a$, which are related to the radiation rate $\gamma_r$ and absorption rate $\gamma_a$ via $\gamma_r = \omega_0 / 2Q_r$ and $\gamma_a = \omega_0 / 2Q_a$, respectively. The $Q_a$ value is related to time-averaged energy absorbed by the cavity during one time periodic, while the $Q_r$ value is defined as the ratio between the time-averaged total energy stored inside the cavity and the time-averaged energy radiated from the cavity during one time periodic.[24]

The parameters $Q_r$ and $Q_a$ determined from the least-squares fitting by fitting Equation (1) to the experimental results, are provided in **Supplementary Note 9**. For the reflective dips from the PMMA nanopillars, it is found that, with the increase in incident angle, the $Q_a$ value increases, while the $Q_r$ value decreases. While for the reflection dip from the middle polariton located F-P cavity, it is found, with an increase in incident angle, the $Q_a$ value increases, and the $Q_r$ value also increases. The increased $Q_a$ values for PMMA cavity and the increased $Q_r$ value for polarion cavity indicate that energy flows from the polaritons to the PMMA, and the periodic PMMA further distributes the polaritonic energy into various designed directions. Therefore, the energy overlap between the PMMA scattering wave and the polaritons creates a coupling channel between these two cavities.

It is also found that the energy flow from polaritons to PMMA is universal for five Bravais lattices. The results for the simple rectangular Bravais lattice are summarized in **Figs. 5d-5f. Fig. 5d** gives the experimental results of angle-resolved reflectance for a simple rectangular Bravais lattice based on PMMA nanopillars supported by a monolayer $MoS_2$/$SiO_2$(300 nm)/Si substrate. **Figs. 5e** and **5f** show experimentally measured angle-resolved reflectance spectra at different incident angles. A vertical shift is introduced in **Fig. 5f** for clarity. Regardless of lattice symmetry, the coupling between PMMA and polaritions promotes the flow of polaritonic energy into PMMA, and the periodic PMMA further distributes the polaritionic energy in different directions in reciprocal space. Thus, the PMMA periodic structure provides an indirect method for tuning the polariton dispersions with specific energy and momentum.

3. Conclusion

In this work, three polariton dispersion branches are realized by coupling excitons in monolayer $MoS_2$ with photons in a $SiO_2$/Si microcavity. It is further demonstrated that the polariton dispersion branches can be directionally controlled in reciprocal space by coupling with five PMMA Bravais lattices, whose periodic structures make the polariton dispersion to shift with different reciprocal vectors under momentum conservation, as observed through angle-resolved reflectance and k-space reflectivity image measurements. Therefore, the flexible creation and manipulation of exciton polaritonic states with specific energy and momentum are realized. The tunable polaritonic dispersion enables the coupling light at different wavelengths at the same incident angle and allows control over light reflectance at arbitrary energy and incident angle, which is significant for designing optical devices with specific properties. The unique ability to control exciton-polariton dispersion will promote its applications in photonics and quantum technologies.

4. Method and experimental section

**Fabrication of 2D Bravais lattices**: Commercial $SiO_2$(300 nm)/Si is commonly used as the sample substrate, and monolayer $MoS_2$ is grown on the $SiO_2$/Si substrate using the CVD method. A positive electron beam resist, PMMA (950 A4 from Resemi), is applied to fabricate the Braavais lattices. First, the PMMA photoresist solution is spin coated on the sample at 500 rpm for 10 seconds, followed by an acceleration to 4,000 rpm at 2,000 rpm s$^{-1}$ for a total time of 1 minute. After spin coating, the PMMA-coated monolayer $MoS_2$ is placed on a hot plate at 180 °C and baked for 5 minutes to cure the PMMA layer. An EBL system (eLINE Plus, Raith) is used to fabricate a 2D Bravais lattice array mask pattern on the PMMA layer. The structure diagrams of five 2D Bravais lattice structures are drawn inversely in Klayout software and loaded into eLINE SEM software. The exposure window is set to 200 μm×200 μm, with a diaphragm size of 7 μm. The auto-reading beam is set to approximately 0.1584 nA, with a corresponding step size of 0.008 μm, and the measurement is set to 200 μC/cm$^2$ for the PMMA 950 A4 resist. Finally, the PMMA-coated sample is immersed in the developer for 2 minutes (MIBK: IPA=1: 3), then soaked in a fixing solution for 40 seconds (pure IPA solution), followed by drying using $N_2$ gas. This process covers the entire electron beam exposure region of the PMMA layer, and the overall size of the manufactured sample is ~200 μm×200 μm.

**Surface morphology characterization**: The morphology of 2D PMMA Bravais lattice is examined using SEM, and the height of the nanopillar is measured using AFM. The AFM measurements employed in the experiment are conducted using a commercially available AFM system (Bruker Dimension XR). All AFM tests are performed under atmospheric conditions at room temperature, utilizing the Scanasyst in Air mode for measurements. The probe selected is the SCANASYST-AIR, with a resonance frequency of 70 kHz and a spring constant of 0.4 N/m. During testing, a scan range of 5 μm×5 μm is used, with a resolution of 256×256 (256 samples per line and 256 lines). The amplitude set point is 250 mV, the drive amplitude is 100 mV, and the scan rate is 0.501 Hz.

**Optical characterization**: Using focal plane Fourier Transform technology, based on the Olympus microscope (IX73), the incident light is illuminated from a halogen lamp and focused on the sample through an objective lens (100×, numerical aperture NA=0.9). The angle-resolved reflectance of the 2D Bravais lattice of PMMA nanopillar supported by $MoS_2/SiO_2$(300 nm)/Si substrate was studied using the NOVA-EX spectrometer (Shanghai Fu Xiang Ideaoptics). In the angular resolution reflectance spectral measurement, one axis of the 2D CCD is used to resolve the angle of the reflected light, while the other axis is used to resolve the wavelength. The in-plane wave vector of the incident light is defined as $\mathbf{k}_\parallel/\mathbf{k}_0=\sin(\theta)$, where $\mathbf{k}_0=2\pi/\lambda$, and $\theta$ is the angle relative to the sample normal. The measured incidence angle ($\mathbf{k}_\parallel/\mathbf{k}_0$) ranges from -0.94 to 0.94 (step size is 0.0023). The reflectance of the cavity structure is measured in the wavelength range of 400 ~ 800 nm with a step size of ~0.75 nm. Background, reference, and signal images are recorded by taking counts on the camera when no light is incident, when light is reflected off a flat Ag on the sample, and when light is reflected off a pattern of interest, respectively. The formula for calculating the final reflectivity image is: reflectance(%)=100%×(signal−background)/(reference−background).

The k-space image measuring device is described in **Supplementary Note 10**. The working principle is that the information carried by the rear focal plane of the objective mirror corresponds to the k-space (Fourier space) information of the sample radiation field. A series of convex lenses are used to image the rear focal plane of the same objective onto the entrance slit of the imaging spectrometer. By obtaining the background image, reference image, and signal image, the reflectance measurement of the dispersed k-space image is obtained. For k-space images, a narrow band-pass filter (Delta Optical Thin Film A/S) with a center of 633 nm and a half-peak full width (FWHM) of 10 nm is placed in the excitation path. The integration time is 400 ms, and the field of view selection range is 50 μm×50 μm. The grating of the Visible Micro Spectrometer (NOVA2S-EX) operates at order 0, with the entrance slit fully open. After the spectrometer, the information in k-space is imaged by a CCD (PIXIS 400).

**Bravais lattice simulation**: The theoretical angular resolution reflectance of the Bravais lattice is simulated using FDTD software. Periodic (Bloch) boundary conditions are applied in the x and y directions, while perfectly matched layers are applied in the z direction. Parallel x-polarized beams of light are incident at different angles on the functional plane of the Bravais lattice. The angular resolution reflectance is obtained by simulating the reflected light from the functional surface.

**Acknowledgements**

L. Sun acknowledges the support from the Beijing Natural Science Foundation (Grant No. Z210006), National Natural Science Foundation of China (Grant No. 12104051), and National Key R&D Plan (2022YFA1405600).

**Author contributions**

Y.W., L.S. and Z.S. conceived the idea, supervised the project, and revised the manuscript. Z.L. and R.B. fabricated the 2D rhombic Bravais lattice samples. Z.L., R.B.,

D.Y., and T.Y. performed the experimental measurements and Z.L., Y.W. and L.S. analysed it. Z.L. and Y.W. performed the theoretical modelling. L.S. and W.L. provided discussions in the simulation. Z.L. and Y.W. performed data analyzation and manuscript preparation. All authors contributed to the discussion of the results.

**Conflict of Interest**

The authors declare no competing interests.

**Data Availability Statement**

The data within this paper are available from the corresponding author upon reasonable request.